# A SURVEY ON AUTHENTICATION AND KEY AGREEMENT PROTOCOLS IN HETEROGENEOUS NETWORKS

Mahdi Aiash[1], Glenford Mapp and Aboubaker Lasebae

[1]School of Science and Technology, Middlesex University, London, UK
`M.Aiash,G.Mapp, A.Lasebae@mdx.ac.uk`

## ABSTRACT

*Unlike current closed systems such as $2^{nd}$ and $3^{rd}$ generations where the core network is controlled by a sole network operator, multiple network operators will coexist and manage the core network in Next Generation Networks (NGNs). This open architecture and the collaboration between different network operators will support ubiquitous connectivity and thus enhances users' experience. However, this brings to the fore certain security issues which must be addressed, the most important of which is the initial Authentication and Key Agreement (AKA) to identify and authorize mobile nodes on these various networks. This paper looks at how existing research efforts the HOKEY WG, Mobile Ethernet and 3GPP frameworks respond to this new environment and provide security mechanisms. The analysis shows that most of the research had realized the openness of the core network and tried to deal with it using different methods. These methods will be extensively analysed in order to highlight their strengths and weaknesses.*

## 1. INTRODUCTION

Future networks is a convergence of different access networks controlled by multiple operators such as 2G/3G, WiMax and the Long Term Evolution (LTE) [1] [2] as the newest wireless technologies being developed and deployed. However with the wide-scale deployment of wireless networks as end-systems, there will now be significant differences in network characteristics in terms of bandwidth, latency, packet loss and error characteristics. These developments mean that, soon it will not be possible to think of the Internet as a single unified infrastructure. It would be better to view the Internet as comprising of a fast core network with slower peripheral networks attached around the core. The core network will consist of a super-fast backbone using optical switches and fast access networks which is mainly based on wired technologies such as Multiprotocol Label Switching (MPLS). Due to the fact that, the connectivity in the peripheral networks will be based on a wide variety of wireless technologies, provided by different operators, various network operators need to cooperate and coexist in the core network.

Unlike current communication systems such as 2G and 3G, which introduce closed environments where the core network is controlled and owned by sole network operator and thus its security is mainly based on the assumption that, the core network is physically secure, the above discussion highlights the fact that we are moving towards an open, heterogeneous environment where the core network is not controlled by a single operator, so multiple operators will have to cooperate. Furthermore, in this environment, new networks providers might choose to join the network and share the spectrum. This introduces dynamic network architecture in contrast to the static current architecture.

This new open and dynamic architecture will bring about new security threats such as initially authenticating the mobile nodes in this environment as well as in the case of handover. The latter issue has been under investigation by different research groups such [3] [4] [5] [6] [7] [8], [9]. However, few research efforts such as the [6] [10] [11] [3] have considered the initial authentication of the Mobile Terminal (MT) in heterogeneous environments.

Most of the previous work realized the openness and dynamic nature of the future networks and have been trying to deal with these issues using different methods. In [11], the authors presumed to have the UMTS infrastructure as a backbone of the core network, while different networks such as WLAN and WiMax could be attached to it. Obviously, this solution does not go along with the open architecture of future networks. Other works such as [3] proposed to use a common platform such as the Extensible Authentication Protocol (EAP) [12] to run their security mechanisms on top of it and thus hiding the difference between different operators. Other solutions have been proposed by the Mobile Ethernet Group in [6] and the Zhing et al, [10] that attempted to address the previous drawbacks by adopting a generic network structure, which is close to an open architecture, and by introducing novel protocols. Due to the fact that, the security protocol in [6] adopts a generic network structure, this protocol will be analysed and verified using formal methods approach. The verification results discovered some security breaches in the deployment of the Mobile Ethernet's Authentication and Key Agreement (AKA) protocol, which highlight the need for a new AKA protocol.

Communication Sequential Processes (CSP) [13] is a formal language to describe the interaction and states in concurrent systems, it has been used to model communicating and security protocols as in [14] and [15]. To verify the CSP models, model checkers such as the Failure Divergence Refinement (FDR) are used. Although modelling and verifying security protocols using CSP and FDR have proven to be effective and widely deployed, modelling directly in CSP is a time-consuming and error-prone. Therefore, a new compiler for generating the CSP description of the protocol was designed by Lowe in [16]. The new compiler is called Casper and it accepts an abstract description of a system and translates it into CSP. This paper will model the security properties of the proposed protocols using Casper and analyse the CSP output with FDR.

The contribution of this work is as follows: Firstly, analysing a number of the AKA protocols for heterogeneous networks namely, the ones proposed by the Zhing et al, the HOKEY, Mobile Ethernet and 3GPP projects. Secondly, using Casper/FDR, we formally model and analyse the initial AKA protocol of the Mobile Ethernet [6]. Thirdly, we analyse the attacks found by Casper/FDR, and verify the security properties of the protocol. Fourthly, based on the performed analysis, we highlight the main source of security threats and propose some recommendations to address them.

The rest of this paper is organized as follows: Section 2 describes the open architecture of the future, heterogeneous networks in terms of its operational components as well as the QoS signalling models as introduced by [17]. The section also describes some related work to address the initial authentication in this environment. Since the Mobile Ethernet framework considers open network architecture, Section 3 explains the initial AKA protocol of the Mobile Ethernet [6] and verifies the protocol using Casper/FDR. The verification results highlights the need for a new AKA protocol. Section 4 explains and formally verifies the three refinement stages, which led to the final version of the protocol. The paper concludes in Section 5.

## 2. OVERVIEW OF FUTURE NETWORKS

In Next Generation Networks, multiple operators have to cooperate in order to provide continuous connectivity. However, since each network operator uses different network architecture, interoperability might be a key challenge. One proposed solution for this problem is having a central management entity to control the resource of the different networks and coordinate the multiple operators. In this regard, the concept of a central management entity was recommended by the ITU-T recommendation [18] for Next Generation Networks (NGNs). The recommendation proposes the concept of the Regulatory Authority as central management entity which controls different network operators and service providers.

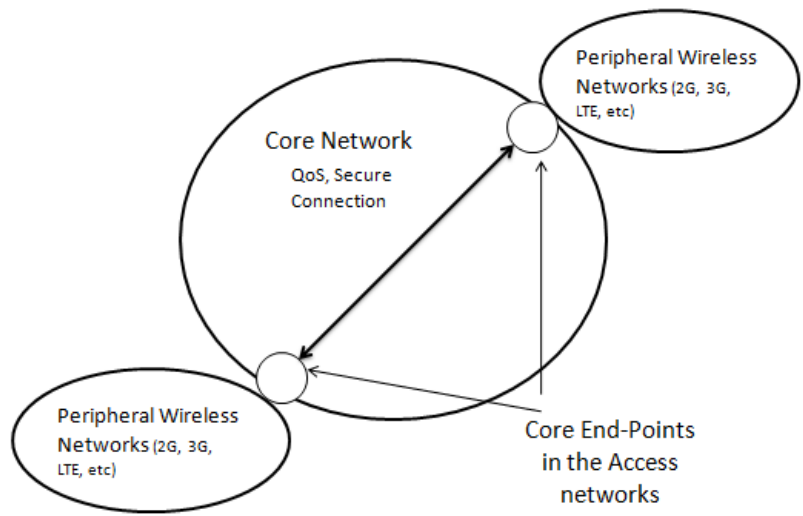

Figure 1. The Architecture of Future Internet

This concept of a central management entity was adopted and enhanced by the Y-Comm group [19] and Daidalos II [20] which introduced the concept of the Core End-Point (CEP) in [17]. As shown in Fig 1, the future Internet could be viewed as composed of several Core End-Points, interconnected over the super-fast backbone of the Internet. Each CEP is responsible for managing multiple, wireless peripheral networks such as Wimax, WiFi or mobile technologies in a local context.

## 3. AKA PROTOCOLS IN HETEROGENEOUS NETWORKS

This section describes some of the related work towards introducing AKA protocols for the initial registration in heterogeneous environments.

### 3.1. AKA and Authorization Scheme Based on Trusted Mobile Platform

The work in [10] has introduced an AKA and authorization scheme to achieve mutual authentication between the user, Mobile Terminal and the SIM card. This scheme deploys passwords in combination with biometric information and Public key Infrastructure (PKI), the scheme also benefits from the Trusted Mobile Platform (TMP) [21] to guarantee the internal integrity of the mobile device. As explained in [10], the proposed scheme achieves many security features such as mutual authentication, protection on wired links as well as resistant to replay and man-in-middle attacks. However, the main drawbacks of the scheme are as follow:

International Journal of Network Security & Its Applications (IJNSA), Vol.4, No.4, July 2012

- The scheme proposes using the PKI. However, this comes at the cost of a higher overhead especially in terms of key management and cryptographic operations [22] [30].

- Many security features of the scheme are based mainly on the hardware architecture of the trusted mobile platform; this implies that the proposed scheme is not generic and might not be compatible with none TMP-supported devices.

### 3.2. AKA Protocol of the Handover Key Working Group (HOKEY WG)

The Internet Engineering Task Force (IETF) handover keying working group (HOKEY WG) [3] is currently developing solutions to provide a secure, media-independent handover, also called inter-technology handover. The solutions are applicable to wireless access technologies based on the Extensible Authentication Protocol (EAP) [12] which is an authentication framework that supports multiple authentication protocols, these are referred to as EAP methods. Regardless of the method, the EAP key hierarchy derives two keys: the Master Session Key (MSK) and the Extended MSK (EMSK) which are used by different methods to derive further keys.

Based on EAP's terminology, three entities are defined: The EAP peer which is the client asking for authentication using an EAP method, the EAP Server which is an entity that terminates the EAP authentication method with the peer; the EAP servers are often, but not necessarily, co-located with Authentication, Authorization and Accounting (AAA) servers. And finally, the EAP authenticator which is the network Access Point that supports the authentication functionality and enforces access control based on the authentication result.

When a mobile terminal (MT) moves between different authenticators, it is desirable to avoid a full EAP authentication to support fast handover. Therefore, the HOKEY group proposed a new method for the EAP known as EAP Re-Authentication Protocol (ERP) [23] which will be discussed in the following sections.

This group is concerned with providing a set of protocols and mechanisms to secure handover. It has introduced an abstract mechanism for delivering root keys from an Extensible Authentication Protocol EAP [12] server to another network server that requires the keys for offering security protected services, such as re- authenticating the EAP-supporting peer using the EAP Re-authentication Protocol (ERP) [23]. The ERP protocol mainly considers the case of handover, and recommends full EAP for initial authentication. However, in either case, the solution is based on the assumption that all access networks support the EAP framework, this assumption might not be feasible in heterogeneous networks since the EAP severs might belong to different operators.

### 3.2.2. An Overview of the ERP Protocol

The ERP is a new extension to EAP to support an EAP method-independent protocol for efficient re-authentication between the peer and an EAP re-authentication (ER) server [23]. It is assumed that, the ER server is collocated with an Authentication, Authorization, and Accounting and Cost (A3C) server [24].

Initially, the MT performs a full normal EAP authentication with the A3C server in its home network. As a result of this authentication, the EAP's keys namely, Master Session Key (MSK) and (Extended Master Session Key) EMSK are derived. However, when the MT roams, the ERP extension is used to achieve authentication between the MT and the ERP-Server in the target network instead of performing a full EAP authentication. This process is referred to as pre-authentication because the keying materials will be launched in the target network before the MT actually joins and it comprises:

For the MT to use the ERP protocol with the access point in the target network, it needs to derive a new re-authentication root key, this key is derived using the EMSK and the domain name of the target network and hence, is called the Domain Specific Root Key (DSRK). Using this key, further domain specific keys such as the DsIK and DSr, MSKs are derived; these will be used to secure the connection between the MT and the network. Additionally, proving the possession of derived keys helps in achieving authentication between the MT and the network.

To provide security without disturbing the handover procedure, the ERP achieves low latency handover by launching the keying materials in the target network before the actual handover takes place. Furthermore, the ERP introduces additional keys shown in Fig 2; these are defined in [23] as follows:

The rRK - re-authentication Root Key, derived from the EMSK.

The rIK - re-authentication Integrity Key, derived from the rRK.

The rMSK - re-authentication MSK. This is a per-authenticator key, derived from the rRK and is delivered to the authenticator.

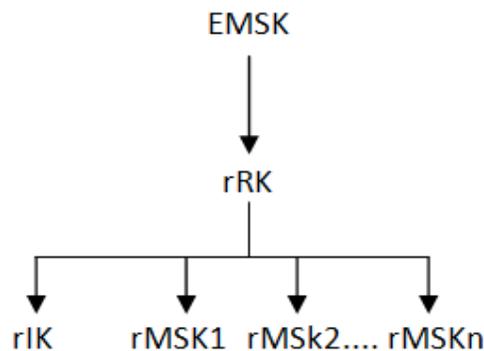

Figure 2. The Key Hierarchy of the ERP Protocol

### 3.2.2. Analysis

The HOKEY's work seemed fairly stable particularly in terms of keys hierarchy and it has influenced the direction of research when developing a Network-Level Authentication and Key Agreement (NL-AKA) protocol. However, the solutions for keys distribution are still being discussed by the HOKEY. Additionally, the ERP extension suffers from some drawbacks which are summarised as follows:

- Although the ERP is based on the EAP platform, it introduces new messages such as EAP-Finish/Re-auth that includes a DSRK and the new domain name. This implies that, all the network entities such as the Access points have to be updated or replaced to support this extra message.

- The EAP-Finish/Re-auth message is sent directly between the MT and the Authenticator in the new network. This message includes the domain name of the target network and is sent in an unprotected manner since there is no security agreement yet between the MT and the target network.

- In the ERP protocol, the Mobile Terminal's home ERP server generates the keys and passes them to the ERP server in the target network. However, in case of heterogeneous environments, this might not be feasible since these ERP severs might belong to different operators.

### 3.3. The Third Generation Partnership Project (3GPP)

The 3GPP group [4] has proposed the integration of 3GPP-WLAN and 3GPP-WiMAX as examples of heterogeneous networks. In both cases, the 3GPP recommends invoking EAP-AKA [25] for the initial authentication. By integrating the 3GPP-AKA [11] protocol and the EAP platform, the EAP-AKA achieves many desired security features such as mutual authentication between the device and the network.

One issue with this approach is that it is fully dependent on specific wireless technology, the 3GPP core network in this case. Whoever wants to add a new wireless access to an existing network will always need to develop a method that integrates wireless access with the 3GPP core infrastructure. Additionally, the solution is based on implementing the EAP platform globally which requires all authentication severs to support the EAP.

### 3.4. Security in the Mobile Ethernet Architecture

Mobile Ethernet Architecture is a Beyond 3G network system for the all IP integrated network using MAC layer technologies [6]. The architecture is based on the Wide Area Ethernet (WAE) which is a virtual private network aimed at providing connectivity based on the Ethernet (MAC) addressing and thus achieves interoperability among different IP-based operators.

For the Network-Level security, Mobile Ethernet has proposed AKA protocols for the initial and handover cases. The proposed protocols consider a generic structure for heterogeneous networks similar to the one in section 2, and since it operates at Layer 2 (L2), it does not require underlying platforms such as EAP and thus, could be used with any operator. Furthermore, as stated in [6], the initial AKA protocol achieves mutual authentication between the mobile terminal and the network and meet many desired security features. Also, this protocol is based on symmetric encryption which makes it less complex to implement for mobile devices. Due to these reasons, this paper will extensively analyse the AKA protocol of the Mobile Ethernet. However, to verify this protocol and make sure it is not vulnerable to security attacks, we use Casper to simulate the protocol and FDR as model checker as detailed in Section 4.2.

#### 3.4.1. Verifying Security Protocols Using Formal Methods and Casper/FDR Tool

Previously, analysing security protocols used to go through two stages. Firstly, modelling the protocol using a theoretical notation or language such as Communication Sequential Processes (CSP) [13]. Secondly, verifying the protocol using a model checker such as Failures-Divergence Refinement (FDR) [26].

However, describing a system or a protocol using CSP is a quite difficult and error-prone task; therefore, Gavin Lowe [16] has developed the CASPER/FDR tool to model security protocols, it accepts a simple and human-friendly input file that describes the system and compiles it into CSP code which is then checked using the FDR model checker. CASPER's input file consists of eight headers as explained in Table 1:

Table 1. The Headers of Casper's Input File

| The Header | Description |
|---|---|
| # Free Variables | Defines the agents, variables and functions in the protocol |
| # Processes | Represents each agent as a process |
| # Protocol Description | Shows all the messages exchanged between the agents |
| # Specification | Specifies the security properties to be checked |
| # Actual Variables | Defines the real variables, in the actual system to be checked |
| # Functions | Defines all the functions used in the protocol |
| # System | Lists the agents participating in the actual system with their parameters instantiated |
| # Intruder Information | Specifies the intruder's knowledge and capabilities |

### 3.4.2. Desired Security Features for AKA protocols

As stated in [27], it is desired for AKA protocols to meet certain security properties. Therefore, a list of these properties will be used to analyse both the initial AKA protocol of [6]

- Mutual Entity Authentication: This is achieved when each party is assured of the identity of the other party.

- Mutual Key Authentication: This is achieved when each party is assured that no other party aside from a specifically identified second party gains access to a particular secret key.

- Mutual Key Confirmation: This requirement means that each party should be assured that the other has possession of a particular secret key.

- Key Freshness: A key is considered fresh if it can be guaranteed to be new and not reused through actions of either an adversary or authorized party.

- Unknown-Key Share Resilience: In this attack the two parties compute the same session key but have different views of their peers in the key exchange. In other words, in this attack an entity A ends up believing that it shares a key with B; although this is the case, B mistakenly believes the key is instead shared with an entity E \= A.

- Key Compromise Impersonation Resilience: This property implies that if the Intruder compromised the long-term key of one party, he should not be able to masquerade to the party as a different party.

## 4. THE INITIAL AKA PROTOCOL FOR THE MOBILE ETHERNET

This section presents a formal analysis of the Initial AKA protocol for Mobile Ethernet proposed by Masahiro et al [6], it deals with providing mutual authentication between the mobile device (M) and the network upon accessing the network for the first time. For this protocol, the security architecture consists of the following network components:

- The Authentication Information Server (AIS): manages the subscriber's information in terms of authentication and authorization.

- The Authentication Server (AS): authenticates the subscribers based on information retrieved from the AIS.

- The Entry Points (EPs): represent one end point for wireless communication and represent Access Points (APs) or Access Routers (ARs).

- The Mobile Device (M): is the mobile terminal accessing the network.

### 4.1. The Protocol Description

The initial AKA protocol of [6] is based on the challenge-response paradigm. By considering the notation in Table 2, the protocol goes as follows:

Initially the mobile device (M) and the AIS pre-share User ID (UID) and user unique key (UUK). When the MD attaches to the access network, it sends its UID and a random number (R1) as a challenge all the way to the AS. The AS appends a freshly created random (R2) to the message and passes it to the AIS. Using the received UID, the AIS looks up in its database and finds the corresponding UUK, then it derives the Master Key (MS) and passes it along with the UID to the AS. The received MS is used by the AS to derive the Authentication Key (AK) and the Secret key (SK), then the AS returns the challenge (R1) encrypted using the AK and a challenge R2 to the mobile device. If the Mobile device managed to derive the required keys, he should be able to verify the received message and compose the response. The AS checks whether the Mobile device possessed the right keys and indicates the end of the authentication process by sending an acknowledgement message.

Table 2. Notations

| The Notation | Description |
| --- | --- |
| M | The Mobile Node |
| AIS | The Authentication Information Server |
| AS | The Authentication Server |
| R1, R2 | Random values |
| E(K, Msg) | Encrypted Msg by key K |
| PRF, PRF2 | Pseudo-random function |
| MS | Master Secret key MS = PRF(UUK, R1 | R2) |
| AK | Authentication Key AK = PRF(MS, R1 | R2) |
| SK | Secret Key used for encryption SK = PRF2(MS, R1 | R2) |

As could be figured out from Fig 3, this version of the protocol might be vulnerable to security threats, which are mainly due to the fact that the derived keys are insecurely distributed to the participating entities. Therefore, the authors in [6], have assumed that, the devices of the architecture are securely installed using mutual authentication and data integrity is maintained in the core network, i.e. between the AIS and the AS. Also, similar to current AKA protocols in current systems such as $2^{nd}$ and $3^{rd}$ Generations [28] [31], it is assumed that the intruder does not know the Key Derivation Functions (KDFs) used to generate the secret and authentication keys. By keeping these assumptions in mind, Casper/FDR tool was used to verify the protocol and find out whether it is still vulnerable to any attacks. A detailed analysis of the protocol is in the following sections.

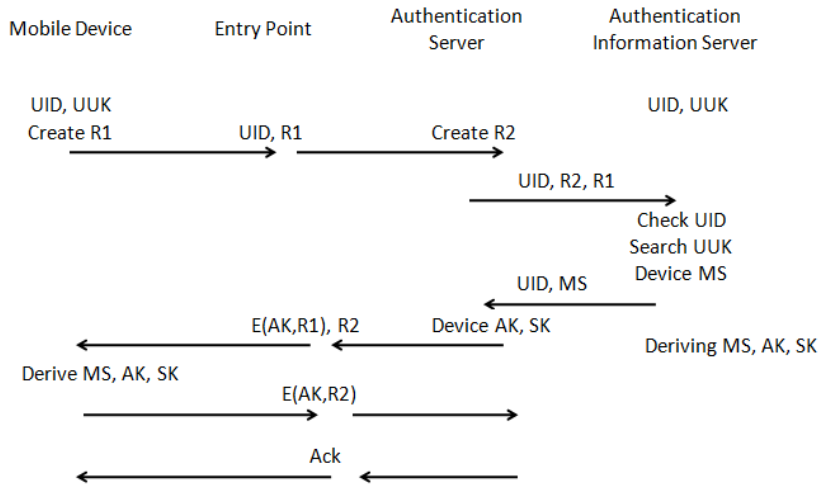

Figure 3. The Initial AKA Protocol of Mobile Ethernet

### 4.2. The Formal Verification of the Mobile Ethernet Protocol

As shown in Fig 3, it is not clear how the Mobile device knows about the Entry Point, this knowledge could not be pre-configured as there is no way to predict which EP the mobile device will use. Similarly, there is a need to justify why the mobile device starts the protocol by sending the the UID,R1 as the mobile device's first message. In order to simulate this interaction in Casper, we introduce the following preliminary messages: the Entry Points' advertisement messages (Adv), The Access Request (AccReq) message, which is used by the Mobile device to indicate its intention to access the network. The Authentication Request (AuthReq) message, sent by the Entry point to trigger the authentication process. None of these messages play a security role; they are only used at the pre-authentication stage, where the entry points advertise their presence.

To formally verify the protocol, a Casper/FDR's input file was prepared. The full input file is given in the Appendix A. However, for conciseness, we only describe the **# Processes**, the **# Specification** and the **# Intruder Information** headings, while the rest are mainly descriptive and of less significance in terms of verifying the protocol.

The **# Protocol Description** section defines the protocol's messages. The notation {m}{k} means that the message (m) is encrypted using the key (k). Also, m%w denotes that the recipient of the message is not supposed to understand the message (m) instead; it should store it in a variable (w) and pass it along to the next recipient. In contrast, the notation w%m means that recipient should be able to encrypt the message (m), stored in the variable (w).

The **# Processes** heading shows that our system comprises four parties: The M represented by the INITIATOR process, the Authenticator process corresponds to the EP; the last two processes namely, the DomainSERVER and CentralSERVER represent the AS and AIS respectively. For each process, the parameters- in the brackets- and variables after the keyword **knows**, define the agents' initial knowledge before running protocol.

The security requirements of the system are defined under the **# Specification** heading. The lines starting with the keyword Secret define the secrecy properties of the protocol. The first line **Secret(M,AK,[AS])** specifies the AK as a secret between the M and the AS. The lines starting with Agreement define the protocol's authenticity properties; for instance **Agreement( M, AS, [R2])** specifies that, the Mobile device is correctly authenticated to the AS and using the random number R2. The Aliveness assertion checks the availability of the participants, e.g. the first Aliveness check **Aliveness (EP, M)** states that when M completes a run of the protocol, apparently with EP, then EP has previously been running the same protocol. Note that EP may have thought he was running the protocol with someone other than M. [16]. A stronger definition of the above Aliveness is specified by the Weak Agreement, for instance **WeakAgreement(EP,M)** assertion insists that M agreed he was running the protocol with EP. So the assertion could be interpreted as follows: if M has completed a run of the protocol with EP, then EP has previously been running the protocol, apparently with M.

```
# Specification
Secret(M,AK,[AS])
Secret(AS,AK,[M])
Secret(M,SK,[AS, EP])
Agreement( M, AS, [R2])
Agreement(AS, M, [AK, R1])
WeakAgreement (EP, M)
WeakAgreement (M, EP)
Aliveness (EP, M)
Aliveness (M, EP)
```

The # Intruder Information heading specifies the Intruder identity, knowledge and capability. The first line identifies the Intruder as Mallory, the Intruder Knowledge defines the Intruder's initial knowledge i.e. we assume the intruder knows the identity of the participants and can generate its own unique key UUK(Mallory). The last two lines specify that all the keys of the Pre-shared keys and Domain specific key are crackable. In other words, the Crackable keyword tells Casper that, the following keys could be compromised by the intruder at any time of the protocol's run.

After compiling the Casper model and feeding the CSP output to FDR, no attacks against the secrecy of the AK and SK keys were found; this is due to the assumption that the Intruder does not know the key derivation functions of these keys despite the fact that the Master Secret Key (MS) is sent unprotected. However, other attack was found against the **Agreement( M, AS, [R2])** and **Aliveness (EP, M)** as shown in Fig 4.

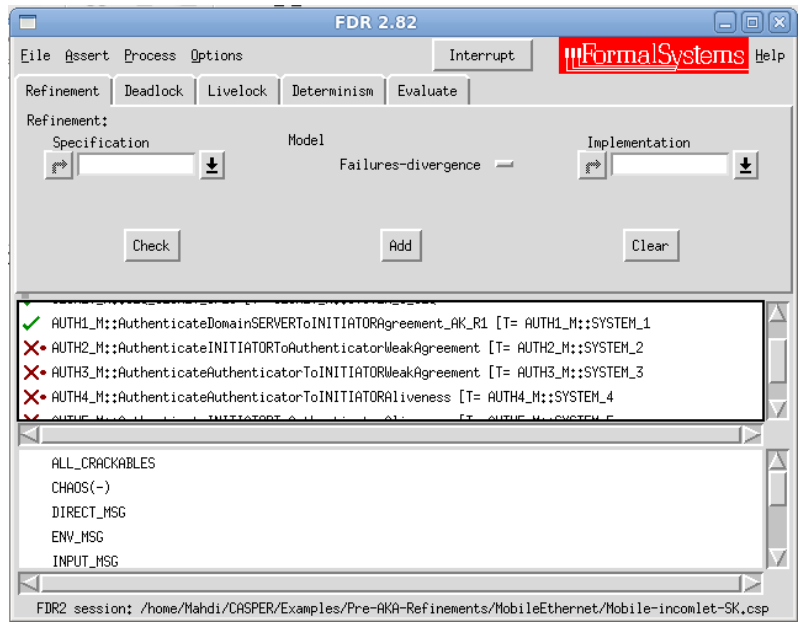

Figure 4. Checking the Mobile Ethernet Protocol using FDR model Checker

We could find the traces for those attacks, which could be translated to the following attack sequence, where The notation I_M for instance represents the intruder taking the Mobile device's identity, either to fake a message (as in the second message 1) or to intercept a message intended for M (as in message 2).

```
 0. -> M : EP, AIS, AS
 1a. M -> I_EP : accReq
 1b. I_M -> EP : accReq
 2a. EP -> I_M : authReq
 2b. I_EP -> M : authReq
 3. M -> I_EP : M, R1
 4. I_EP -> AS : M, R1, h(M, R1)
 5a. AS -> I_AIS : M, R1, R2, h(M, R1, R2)
 5b. I_AS -> AIS : M, R1, R2, h(M, R1, R2)
 6a. AIS -> I_AS : MS, M, h(MS, M)
 6b. I_AIS-> AS : MS, M, h(MS, M)
 7. AS -> I_EP : R2, {R1}{AK}, h(R2, {R1}{AK})
 8. I_EP -> M : {R1}{AK}, R2
 9. M -> I_EP : {R2}{AK}
 10. I_EP -> AS : {R2}{AK}, h({R2}{AK})
 11. AS -> I_AIS : hoackm, h(hoackm)
 12. I_EP -> M : hoackm
```

The discovered attack could be depicted as in Fig 5 and explained as follows:

- Initially, the intruder intercepts and replays the messages between the (M) and the (EP) as in messages 1a, 1b, 2a, 2b. Also, the intruder impersonates the Entry Point (I_EP) to intercept message 3 and fake message 4 towards the Authentication Server (AS).

- Messages 5a, 5b, 6a, 6b between the AS and AIS are intercepted and passively replayed, Eventually, the intruder manages to get the new random (R2) in message 7 and then it impersonates the EP once the new random (R2) to run the protocol as messages 8,9, 11 and 12.

In other words, this attack could be interpreted as follows: The Mobile device (M) thinks he has successfully completed a run of the protocol apparently with EP, while in reality it is with the Intruder, and EP has not previously been running the protocol.

### 4.3. Protocol Analysis and Security Consideration

In this section, we discuss how our formal modelling with Casper allows checking the security requirements described in Section 3.4.2.

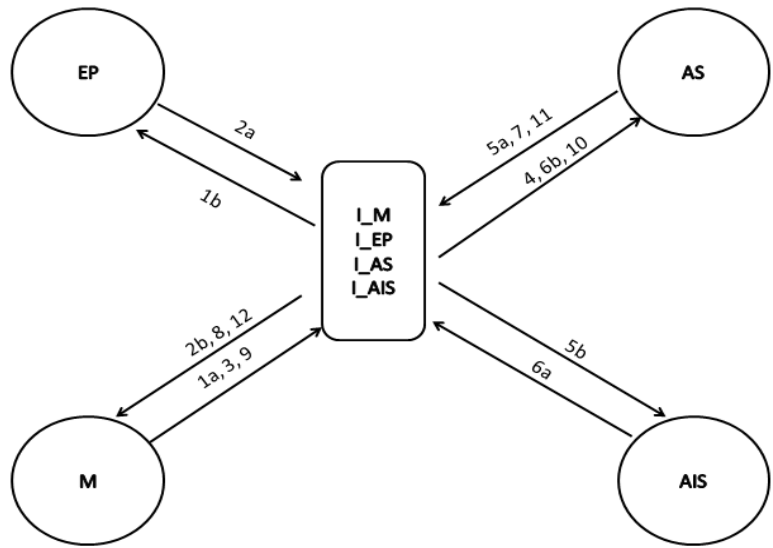

Figure 5. The Discovered Attack

- Mutual Entity Authentication: As stated in [27], entity authentication involves corroboration of a claimant's identity through actual communications with an associated verifier during execution of the protocol itself. Since the protocol does not consider verifying the identity of the participants and based on the discovered attack, we could claim that this protocol could not meet this feature.

- Mutual Key Authentication: the mutual authentication between the M and the AS is based on the secrecy of the AK. We got Casper to check this using the Secret (M, AK, [AS]) and Secret (AS, AK, [M]) assertion checks. Since no attack was found against the key secrecy, this property is met.

- Mutual Key Confirmation: Casper verifies this requirement by using the DECRYPTABLE (m, K) which checks if the message (m) is decryptable by the key (K). We performed a similar check after messages 8 and 10 as shown in the Protocol Description heading to verify that the valid Authentication key (AK) is possessed by the other party. If any of the checks fails the protocol aborts.

International Journal of Network Security & Its Applications (IJNSA), Vol.4, No.4, July 2012

- Key Freshness: This property is guaranteed by including a fresh random value R1, R2 in the key derivation functions of the keys MS, AK and SK.

- Unknown Key Share Resilience: The afore-explained attack implies that the UKS was not met. Despite of the fact that, the mobile device (M) and the AS share the Authentication Key (AK), the M mistakenly believes that the intruder holds this key as well. Casper/FDR indicates this fact by highlighting an attack against the Agreement and Aliveness assertions in the # Specifications header.

- Key Compromise Impersonation Resilience: this property could be modelled by specifying the long-term keys as crackable and then checking the Authenticity assertions. Casper verifies no breach against the authenticity feature

It is obvious from the discussion above that, the initial AKA protocol failed to meet some security requirements, which are mainly related to the discovered authentication attack in Fig 5. Although the protocol presumed the core network entities to be securely installed and the integrity of the exchanged messages to remain intact between the AIS and the AS, the fact that an attack could still be discovered could be due to the Intruder managing to intercept the connections in the core network. This raises the issue of the need for providing a better security in the core network. Initially, the core network has been assumed to be physically secure, this assumption was valid in the closed, homogeneous environments, where the core network was controlled by a sole operator. However, this assumption does not hold in the case of future networks, where the core network represents open, multi-operators environments. Additionally, there is a need to deal with identification-related attacks to meet the Mutual Entity Authentication property.

Furthermore, the process of deriving the keying materials in the Initial AKA protocol of [6] does not define the keys' usability scope. Therefore, there is a need to propose a more stable key hierarchy that specifies the scope of each derived keys.

## 5. CONCLUSION

This paper discussed several research efforts, which have been trying to address the issue of authenticating the mobile nodes when they initially join the heterogeneous environment. The discussion showed that most of the solutions had realized the threats resulting from the open nature of future networks and as a result different approaches were proposed. Some solutions tried to conceal the divergence of the core network either by considering a specific technology as a backbone of the core network, or by deploying a common framework on top of which security protocols could be installed and run. The Mobile Ethernet group proposed a new AKA, which considers an open network architecture. Analysing and verifying the Mobile Ethernet's AKA protocol using Casper/FDR shows that the protocol is vulnerable to an authentication attack. Also, the protocol failed to meet some desired security properties, which could be ascribed to the lack of security in the core network. Hence, this work shows that as the core of the network is opened, more attacks will be possible on network entities that previously were protected in closed environments.

## REFERENCES

[1]     J.H. Schiller."Mobile communications" (2003), 2nd ed. London :Addison-Wesley , pp. 136-154.

[2]     Long Term Evolution Protocol Overview, (2008)  Freescale Semiconductor, http://www.freescale.com/files/wireless_comm/doc/white_paper/LTEPTCLOVWWP.pdf. [Accessed 07 July 2012].

[3]     Internet Engineering Task Force, Handover keying working group (hokeywg) http://www.ietf.org/html.charters/hokey-charter.html. [Accessed 07July 2012].
210

[4]     3rd Generation Partnership Project (3GPP), http://www.3gpp.org/ . [Accessed 07July 2012].

[5]     Institute of Electrical and Electronics Engineers, (2007), IEEE 802.21/D8.0, DraftStandard for Local and Metropolitan Area Networks: Media IndependentHandover Services..

[6]     K Masahiro, Y Mariko, O Ryoji, K Shinsaku, T Tanaka (2004): Secure service andnetwork framework for mobile ethernet. Wirel Personal Commun.29,161– 190.

[7]     3rd Generation Partnership Project, 3GPP Technical Specifications. (2006): 3GSecurity; WLAN interworking security (Release 7)..

[8]     AS Ali, (2010): Authentication and key management in heterogeneous wirelessnetworks, PhD Thesis, Electrical and Computer Engineering (The Universityof British Columbia, 2010).

[9]     M. Aiash, G. Mapp, A. Lasebae, R. Phan and J. Loo,A (2012) Formally Verified AKA Protocol For Vertical Handover in Heterogeneous Environments using Casper/FDR, EURASIPJournal on Wireless Communications and Networking 2012,2012:57 doi:10.1186/1687-1499-2012-57. OpenSpringer.

[10]    Y. Zheng, D. He, X. Tang and H. Wang, (2005): AKA and AuthorizationScheme for 4G Mobile Networks Based on Trusted MobilePlatform, In Proceedings of ICICIS.

[11]    M. Aiash, G. Mapp, A. Lasebae, R. Phan, (2010).Providing Security in 4G Systems: Unveiling theChallenges. In The Sixth Advanced InternationalConference on Telecommunications, AICT 2010 .Barcelona, Spain, 9-15 May 2010.

[12]    B Aboba, L Blunk, J Vollbrecht, J Carlson, H Levkowetz, (2004): Extensible Authentication Protocol (EAP) RFC 3748 .

[13]    P. Ryan, S. Schneider, M. Goldsmith, G. Lowe and A,W. Roscoe, (2010):  The modelling and analysis of security protocols ,PEARSON Ltd.

[14]    S Xu, C Tser Huang, MM Matthews, (2008): Modeling and analysis of IEEE 802.16PKM Protocols using Casper FDR,in Wireless Communication Systems,ISWCS' 08, Reykjavik, Iceland, 653–657.

[15]    M. Aiash, G. Mapp, R. Phan A. Lasebae, J.Loo  (2012)  *A Formally Verified Device Authentication Protocol Using Casper/FDR* TrustComm 2012, LiverPool, UK.

[16]    G Lowe, P Broadfoot, C Dilloway, M Hui, Casper, a compiler for the Analysis of security protocol, http://www.comlab.ox.ac.uk/gavin.lowe/Security/ Casper/. [Accessed 07 July 2012].

[17]    M Aiash, G Mapp, A Lasebae,  (2011): A QoS framework for Heterogeneous Networking, in ICWN2011, London UK, 1765–1769.

[18]    International Telecommunication Union (ITU-T), (2004):  Global InformationInfrastructure, Internet Protocol Aspects And Next Generation Networks,Y.140.1.

[19]    Y-Comm Research, http://www.mdx.ac.uk/research/areas/software/ ycomm_research.aspx. [Accessed 07 Jul. 12].

[20]    S Sargento, V Jesus, F Sousa, F Mitrano, T Strauf, C Schmoll, J Gozdecki, GLemos, M Almeida, D Corujo,(2007): Context-Aware End-to-End QoS Architecturein Multi-technology, Multi-interface Environments, in16 th Mobile and Wireless Communications Summit,Budapest 1–6 .

[21]    Y. Inamura, T. Nakayama, A. Takeshita, Trusted MobilePlatform Technology for Secure Terminals. NTT DoCoMoTechnical Journal, 7, 25-39.

[22]    P. Ebinger M. Jalali-Sohi, (2002): Towards efficient pkis for restricted mobile devices, CCN'02.

[23]    L. Dondeti V. Narayanan,(2008): Eap extensions for eap re-authentication protocol (erp), Standards Track 5296.

[24]    A. Rubens W. Simpson C. Rigney, S. Willens, (2000): Remote authentication dial in user service (radius), RFC 2865, Network Working Group.

## APPENDIX

**A:** Code for Formal Analysis of the Initial AKA Protocol for Mobile Ethernet

# Free Variables

M: MobileTerminal
EP : AccessRouterAuthenticator
AS : DomainA3CServer
AIS : CentralA3CServer
AuthID : Identity
Initauth : Flags
R1 : initialSeq
R2 : Sequence
UUK : MobileTerminal-> PresharedKeys
AK : AuthenticationKeys
SK : SecretKeys
MS: Domainspecifickey
RPF: PresharedKeys x initialSeq -> Domainspecifickey
rpf: initialSeq x Domainspecifickey ->  AuthenticationKeys
F3: initialSeq x Domainspecifickey -> SecretKeys
h : HashFunction
AccReq, AccRes,AuthReq, Adv: Messages
HoAckm : AcknowledgementMessage
InverseKeys = (AK, AK), (UUK, UUK) , (SK, SK),  (MS, MS), (RPF,RPF), (rpf,rpf),(F3,F3)

# Processes
INITIATOR(M, EP, R1,AuthID,Initauth, AccReq, AuthReq) knows UUK(M)
Authenticator(EP,M,AS, AuthReq, Adv,AccRes)
DomainSERVER(AS,AIS, R2, HoAckm)
CentralSERVER(AIS) knows UUK(M)

# Protocol Description

0. -> M : EP, AIS, AS
1. M -> EP: AccReq
2. EP -> M : AuthReq
< MS := RPF(UUK(M), R1);/
AK:= rpf(R1, MS)>
3. M -> EP : M,R1
4. EP -> AS : M,R1, h(M,R1)
5. AS -> AIS : M,R1, R2, h(M,R1,R2)
< MS := RPF(UUK(M), R1)>
6. AIS -> AS : MS,M, h(MS, M)
< AK:= rpf(R1, MS)>
7. AS -> EP: R2,({R1}{AK}%z)%x, h(R2,({R1}{AK}%z)%x)
8. EP -> M : x%({R1}{AK}%z), R2
[decryptable(z, AK)and nth(decrypt(z, AK), 1) == R1]
<SK:= F3(R1, MS)>
9. M -> EP : ({R2}{AK}%y)%q
10. EP -> AS: (q%{R2}{AK})%y, h((q%{R2}{AK})%y)
[ decryptable(y, AK)and nth(decrypt(y,AK), 1) == R2]
<SK:= F3(R1, MS)>
11. AS -> EP :HoAckm, h(HoAckm)
12. EP -> M : HoAckm

# Specification

Secret(M,AK,[AS])
Secret(AS,AK,[M])
Secret(M,SK,[AS, EP])
Agreement( M, AS, [R2])
Agreement(AS, M, [AK, R1])
WeakAgreement (EP, M)
WeakAgreement (M, EP)
Aliveness (EP, M)
Aliveness (M, EP)

# Actual Variables
m, Eve: MobileTerminal
ep : AccessRouterAuthenticator
as : DomainA3CServer
ais : CentralA3CServer
Authid : Identity
InitAuth : Flags
r1 : initialSeq
r2 : Sequence
ak : AuthenticationKeys
sk : SecretKeys
ms: Domainspecifickey
accReq, accRes,authReq, adv: Messages

hoackm : AcknowledgementMessage
InverseKeys = (ms, ms), (ak, ak), (sk, sk)

# Functions
symbolic UUK, RPF, rpf, F3

# System
INITIATOR(m,ep, r1,Authid,InitAuth, accReq, authReq)
Authenticator(ep,m,as, authReq,adv, accRes)
DomainSERVER(as,ais, r2,hoackm)
CentralSERVER(ais)

# Intruder Information
Intruder = Eve
IntruderKnowledge = {m, as, Eve, ais, Authid, ep, UUK(Eve)}
Crackable = PresharedKeys
Crackable = Domainspecifickey

**Authors**

Mahdi Aiash: This author received his Master and PhD degrees from Middlesex University, London, UK. Dr. Aiash is involved in many research efforts such as the Y-Comm research group and the Wireless Sensors group. His main interest is in the area of network and information security, ubiquitous and pervasive communication. He is an Associate member of the IEEE and IEEE ComSoc since 2007 and a Programme Committee in many conferences and Journals.

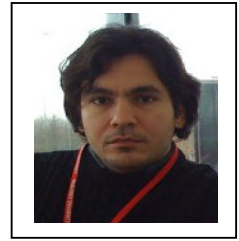

Glenford Mapp: This author received his PhD from the Computer Laboratory; University of Cam-bridge in 1992.He is also a Principal Lecturer in Computer Networks at Middlesex University in North Lon-don and a Visiting Fellow in the LCE Technology Group at the Computer Laboratory, University of Cambridge. He worked on a number of networking-oriented projects and Proposed the X-Windows Teleporting project, which later evolved into Virtual Network Computing, http: //www.realvnc.com.He also led the early stages of the CLAN project, which developed very low latency networking technology for the local area. A new company in Cambridge called Level 5

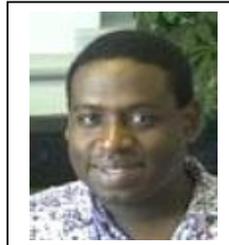

Networkshttp://www.level5networks.comis bringing some exciting new low latency networking products to the marketplace. He is the chief architect of Y-Comm, a new architecture for future mobile communications systems. See :http://www.mdx.ac.uk/ research/areas/software/ycomm research.aspx. He is working, along with his colleagues at Middlesex, on developing performance models for mobile and distributed architectures. Currently he is working on proactive vertical handover algorithms, network memory storage systems, flexible transport protocols and network resilience.

Aboubaker Lasebae: This author got his BASc from the University Regina, Canada and his MSc from the University of Southampton, England and PhD from Middlesex University. Currently, he is a principal lecturer at the School of Engineering and Information Systems and the director of postgraduate programmes for Computer Communications and the programme Leader of Computer and Network Security. He has several publications in the areas of QoS, Network Security, Wireless Networks and Mobile IP.

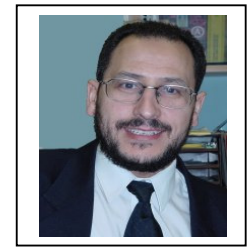